\documentclass[11pt]{article}

\usepackage[final]{acl}

\usepackage{times}
\usepackage{latexsym}
\usepackage[T1]{fontenc}
\usepackage[utf8]{inputenc}
\usepackage{microtype}
\usepackage{inconsolata}
\usepackage{amsmath,amssymb,amsthm,mathtools}
\usepackage{booktabs}
\usepackage{graphicx}
\usepackage[nameinlink,noabbrev]{cleveref}
\usepackage{enumitem}

\newtheorem{theorem}{Theorem}
\newtheorem{proposition}[theorem]{Proposition}
\newtheorem{lemma}[theorem]{Lemma}

\theoremstyle{definition}

\theoremstyle{remark}
\newtheorem{remark}[theorem]{Remark}

\newcommand{\R}{\mathbb{R}}
\newcommand{\cB}{\mathcal{B}}
\newcommand{\one}{\mathbf{1}}
\newcommand{\E}{\mathbb{E}}
\newcommand{\ind}{\mathrm{I}}
\newcommand{\noise}{\mathrm{N}}

\title{Prediction-Market Seed Capital Recovery from Noise-Dominant Flow}
\author{%
  \textbf{Yankai Chen}\textsuperscript{1,2},
  \textbf{Bowei He}\textsuperscript{1,2},
  \textbf{Zhuohan Xie}\textsuperscript{2},
  \textbf{Zhiwei Liu}\textsuperscript{3},
  \textbf{Fan Zhang}\textsuperscript{2},
\\
  \textbf{Xueqing Peng}\textsuperscript{4},
  \textbf{Yan Wang}\textsuperscript{4},
  \textbf{Lingfei Qian}\textsuperscript{4},
  \textbf{Xue Liu}\textsuperscript{1,2}
\\
  \textsuperscript{1}McGill University \quad
  \textsuperscript{2}MBZUAI \\
  \textsuperscript{3}The University of Manchester \quad
  \textsuperscript{4}The Fin AI
}

\hypersetup{
  pdfauthor={Yankai Chen, Bowei He, Zhuohan Xie, Zhiwei Liu, Fan Zhang, Xueqing Peng, Yan Wang, Lingfei Qian, Xue Liu},
  pdftitle={Prediction-Market Seed Capital Recovery from Noise-Dominant Flow},
  pdfsubject={FinNLP Workshop @ EMNLP 2026}
}

\newcommand{\newtext}[1]{#1}

\begin{document}
\maketitle

\begin{abstract}
Automated prediction markets require sponsors to prefund liquidity before
observing order flow, creating a financing challenge at launch. We study whether
nonnegative charges conditioned on observable payoff direction can improve
recovery of this prefunded capital while limiting their effect on informed
participation. We develop Seed Capital Flow (SCF), a direction-conditioned levy,
in a stylized binary cost-function market with informed and
liquidity-motivated traders. When order composition differs across directions,
SCF concentrates the permitted fee burden on the direction with relatively more
liquidity-motivated flow, whereas a uniform fee spreads it across both
directions. Under a sufficiently tight common retention constraint, this
allocation yields higher expected recovery capacity and can make additional
liquidity choices financially viable. Synthetic numerical audits examine
robustness to alternative flow patterns, stochastic arrivals, and label misspecification. The
results characterize a mechanism-design tradeoff rather than an empirical
prediction: the market remains prefunded, recovery is expected rather than
guaranteed, and the analysis is limited to an opening-cohort setting.
\end{abstract}

\section{Introduction}
\label{sec:introduction}

\newtext{Automated prediction markets can aggregate the probabilistic signals produced
by financial NLP systems
\citep{xie-etal-2026-finchain,elbadry-etal-2026-sahm,zhou-etal-2026-fincards,wang2026finauditing},
but such a market must first exist, and here they face a cold-start problem.}
Deeper liquidity makes
a cost-function market easier to trade against, but the sponsor must commit a
larger worst-case-loss budget before observing any order flow. No fee can create
revenue when no potential trader arrives. We instead ask whether fee design can
reduce the traffic required to recover prefunded seed capital in expectation.

Uniform fees earn revenue from liquidity-motivated orders but also deter orders
that convey information\newtext{, such as those placed by forecasting models and
trading agents \citep{lin2025cspo,peng2026herculean,gong2026shijianbench}}. This tension changes when payoff direction predicts
order composition. If one side has a higher liquidity-to-informed ratio, a
uniform fee spends the informed-participation budget equally across two unequal
directions. We call our policy Seed Capital Flow (SCF), a direction-conditioned
recovery levy. It gives fee relief to the candidate-information-dominant direction and
concentrates the permitted burden
on the candidate-noise-dominant direction. This is a Ramsey-style allocation of
a common screening budget: the platform observes direction, not trader type.

We analyze this idea in a binary logarithmic market scoring rule with fixed-size
orders and one opening cohort. The same primitives determine three objects: the
exact finite-trade cost faced by an arriving trader, the participation of
informed and liquidity-motivated traders, and the sponsor's prefunded
worst-case-loss budget. Raw surplus distributions and type--direction arrival
intensities are fixed before liquidity is chosen. Glosten--Milgrom-style
primitives make informed candidate flow symmetric under a common prior, while
liquidity-motivated flow is more concentrated in one direction. SCF and the
uniform benchmark face the same nonnegativity, fee-cap, and retention rules.

\paragraph{Result.}
Under a tight retention allowance, the unique SCF optimum exempts the
candidate-information-dominant direction and charges the
candidate-noise-dominant direction. The optimal uniform policy splits the same
permitted burden between both directions. Although the two policies deter the
same expected number of informed orders, SCF earns strictly
more expected levy revenue at every admissible liquidity level.

The pointwise revenue gap changes the launch boundary. Each mechanism has a
minimum potential-arrival scale at which at least one positive liquidity level
can recover its seed capital in expectation, and the SCF threshold is strictly
lower. Between the two thresholds, some positive liquidity levels are
recoverable with SCF, while none are recoverable with any
feasible uniform levy. Above the uniform threshold, every uniformly recoverable
level is recoverable under SCF, and the inclusion is strict. The separation
also survives an informed-order-loss allowance fixed across candidate liquidity
levels once the platform imposes a positive minimum depth.

The contribution is deliberately narrow. We identify a sufficient combination
of fixed raw Glosten--Milgrom-style flow, a common informed-order-loss rule, and
directional asymmetry that makes the weak menu-inclusion comparison strict at
every admissible depth. Exact LMSR geometry then converts that strict screening
gain into an arrival-threshold and recoverable-set result. Differentiated fees,
weak dominance from a larger policy menu, and noise-flow financing are not
claimed as new by themselves. A structured simulation audit maps the size of the
threshold effect, supplies a wrong-direction negative control, quantifies exact
finite-cohort chance thresholds, and probes nonuniform surplus distributions
outside the theorem.

\section{Related Work}
\label{sec:related}

Cost-function prediction markets provide continuous liquidity with bounded
sponsor loss \citep{hanson2003combinatorial,chen2007utility,abernethy2013efficient}.
Profit-charging and liquidity-sensitive market makers use revenue or activity
to offset subsidy and adjust liquidity
\citep{othman2012profit,othman2013practical,li2013adaptive}; recent work studies
liquidity pools, general provisioning interfaces, online adaptation, dynamic
loss allocation, and quadratic fees
\citep{amini2026decentralized,bhaskara2026liquidity,nueve2026adaptive,moallemi2026uniform,nueve2025smooth}.
We instead fix LMSR depth and compare recovery policies under a common
informed-order-retention rule.

Our adverse-selection primitives follow the insight that direction changes a
market maker's posterior when informed and uninformed traders coexist
\citep{glosten1985bid}. Related models connect those primitives to automated
market-maker curves or optimize prediction-market bid and ask quotes
\citep{nadkarni2024adaptive,feil2026optimal}. Our one-cohort model has no belief
updating, spread choice, or inventory control.

Directional and information-sensitive AMM fees are also established.
Prior work studies asymmetric or dynamic charges for toxic arbitrage and noise
flow, nonconstant-fee accuracy--loss frontiers, and uninformed-flow financing
of equilibrium liquidity
\citep{alexander2024fees,baggiani2026dynamic,rao2023triangle,adams2025amamm}.
The uniform menu is nested in the SCF policy class, so weak dominance is
generic. Our strict result instead uses unequal direction-level
liquidity-to-informed ratios, a binding common retention constraint, and the
exact LMSR seed requirement. Payoff-representation invariance and split-proof
fee design are treated as implementation constraints, not standalone novelty
\citep{frongillo2018axiomatic,frongillo2024equivalence,bichuch2026axioms}.

\newtext{Financial NLP benchmarks supply the forecasts such markets aggregate
\citep{wang2026convfinre,zhang-etal-2026-finreporting,xie2026cleffinmmeval,xie2026finmmevaltask1,xie2026finmmevaltask2},
and autonomous agents form a growing share of traders
\citep{xie2026finmmevaltask3,safder2026finpersona}.
A companion mechanism adapts fees to trade span for sequential aggregation
\citep{chen2026spanpm}; we instead study financing the launch.}

More generally, price-discrimination results already show that a discriminatory
menu weakly dominates uniform pricing and characterize when profit gaps can be
large \citep{malueg2006bounding,bergemann2022third}. Our claim is correspondingly
bounded: among the public primary sources reviewed through August 11, 2026, we
did not identify the same fixed-raw opening-cohort primitives combined with a
common informed-order-loss allowance and a strict LMSR arrival-threshold
separation. This scoped search statement is not a priority or firstness claim.

\section{Market and Opening-Cohort Model}
\label{sec:model}

\subsection{Binary payoff space and LMSR capital}

There are two mutually exclusive outcomes, \(Y\) and \(N\).
A contingent claim is a vector \(r=(r_Y,r_N)\in\R^2\), and \(\one=(1,1)\).
Portfolios \(r\) and \(r+\alpha\one\) differ only by the sure payoff \(\alpha\).
The risky quotient therefore has one coordinate,
\begin{equation}
    x(r)=r_Y-r_N.
    \label{eq:risky-coordinate}
\end{equation}
The sign of \(x(r)\) defines the economic payoff direction, independently of the order interface.
For example, \((z,0)\) (buy \(z\) YES) and \((0,-z)\) (sell \(z\) NO) share risky coordinate \(z\) and differ only by \(z\one\).

For \(b>0\), the binary LMSR cost function is
\begin{equation}
    C_b(q)=b\log\!\left(e^{q_Y/b}+e^{q_N/b}\right).
    \label{eq:lmsr-cost}
\end{equation}
We initialize at \(q^0=(0,0)\), so the price is \((1/2,1/2)\).
The cost is translation invariant:
\begin{equation}
    C_b(q+\alpha\one)=C_b(q)+\alpha.
    \label{eq:lmsr-translation}
\end{equation}
Its worst-case loss from the initial state is
\begin{equation}
    K(b)=b\log 2.
    \label{eq:seed}
\end{equation}
The outside option \(b=0\) means that the platform does not launch the product.

Fix an indivisible risky lot \(z>0\).
The canonical zero-sum representatives are
\begin{equation}
\begin{aligned}
    r_h=\left(\frac z2,-\frac z2\right),
    \qquad
    r_n=-r_h.
\end{aligned}
    \label{eq:canonical-lots}
\end{equation}
At the initial state, the exact LMSR cost of either representative is
\begin{equation}
    D(b)
    :=C_b(q^0+r_h)-C_b(q^0)
    =b\log\cosh\!\left(\frac{z}{2b}\right).
    \label{eq:D}
\end{equation}
Thus both directions have the same finite-trade cost; the benchmark removes LMSR-side asymmetry.

\begin{lemma}[LMSR geometry]
\label{lem:lmsr-geometry}
The function \(D\) is continuous and strictly decreasing on \((0,\infty)\), with
\begin{equation}
\begin{aligned}
    \lim_{b\downarrow0}D(b)=\frac z2,
    \qquad
    \lim_{b\to\infty}D(b)=0.
\end{aligned}
    \label{eq:D-limits}
\end{equation}
Consequently, for every \(w_N\in(0,z/2)\), there is a unique \(b_0>0\) such that \(D(b_0)=w_N\).
\end{lemma}

\subsection{Timing and the recovery object}

The platform proceeds in two stages.
At time zero it chooses \(b>0\), posts the full capital in \cref{eq:seed}, and announces a levy policy.
During a specified recovery window, potential fixed-lot orders arrive.
The platform calls \(b\) \emph{recoverable in expectation} if expected levy receipts during that window are at least \(K(b)\).
The initial capital remains necessary even when this inequality holds.

Our closed form is a fixed-opening-state cohort model.
Participation is evaluated against the initial LMSR schedule in \cref{eq:D}; the model does not treat \cref{eq:D} as the exact cost faced by every order along an endogenous multi-order state path.
The cohort may be viewed as a launch-screening benchmark or as the reduced form of an opening batch.
A sequential implementation would have to replace the demand functions below with expectations over the evolving state.

\subsection{Latent types and fixed raw primitives}

Potential orders have latent type \(T\in\{\ind,\noise\}\) and observable payoff direction \(s\in\{h,n\}\).
Type \(\ind\) is informed; type \(\noise\) has an outcome-independent liquidity motive.
The labels \(h\) and \(n\) will denote the candidate-information-dominant and candidate-noise-dominant directions, respectively.
The platform observes \(s\), not \(T\).

Direction is a policy-invariant attribute of the candidate order. Conditional
on \((T,s,W)\), the announced levy changes only the execute-versus-abstain
decision; a candidate cannot substitute an \(h\) order for an \(n\) order or
vice versa. Thus the type--direction intensities below are primitives rather
than equilibrium responses to the fee menu. This restriction is material and
is revisited in \cref{sec:limitations}.

Let \(W\) be an order's raw linear advantage before subtracting the LMSR finite-trade cost and the levy.
Fix type-specific supports
\begin{equation}
    0<w_N<w_I\le \frac z2,
    \qquad
    d:=w_I-w_N>0.
    \label{eq:supports}
\end{equation}
For each type--direction pair, fix a density \(c_{T,s}>0\), independent of \(b\).
At arrival scale \(m>0\), the raw expected intensity is
\begin{equation}
    m c_{T,s}\,dW
    \quad\text{for }W\in[0,w_T],
    \label{eq:raw-intensity}
\end{equation}
and zero outside the support.
These intensities need not sum to one.

For an informed order, direction \(h\) can be represented by posterior
\(\mu=1/2+W/z\), and direction \(n\) by \(\mu=1/2-W/z\).
Condition \cref{eq:supports} keeps both posteriors in \([0,1]\).
For a liquidity-motivated order, \(W\) is a private trading benefit independent of the realized outcome.

Use the density notation
\begin{equation}
\begin{aligned}
    i_s=c_{\ind,s},
    \qquad
    n_s=c_{\noise,s},
    \qquad
    c_s=i_s+n_s.
\end{aligned}
    \label{eq:density-notation}
\end{equation}
We impose
\begin{equation}
    i_h=i_n=:i>0,
    \qquad
    n_n>n_h>0.
    \label{eq:flow-asymmetry}
\end{equation}
The first restriction gives a symmetric common-prior informed construction.
The second places more liquidity-motivated raw intensity in direction \(n\).
The strict comparison comes from this flow asymmetry, not from directional LMSR curvature.

For \(b>b_0\), define the fee headroom
\begin{equation}
    a(b):=w_N-D(b)>0.
    \label{eq:headroom}
\end{equation}
The corresponding informed-trader headroom is \(a(b)+d\).
If direction \(s\) faces a per-lot levy \(\ell_s\in[0,a(b)]\), an order executes exactly when \(W\ge D(b)+\ell_s\).
The resulting demands are
\begin{equation}
    Q_{\ind,s}(\ell_s;b)
    =m i\,[a(b)+d-\ell_s].
    \label{eq:QI}
\end{equation}
\begin{equation}
    Q_{\noise,s}(\ell_s;b)
    =m n_s\,[a(b)-\ell_s].
    \label{eq:QN}
\end{equation}
All four type--direction cells remain in their linear regions on the stated levy domain.

At zero levy, the informed share among candidate orders in direction \(s\) is
\begin{equation}
    \rho_s(b)
    =
    \frac{i[a(b)+d]}
         {i[a(b)+d]+n_sa(b)}.
    \label{eq:candidate-posterior}
\end{equation}
Hence \(\rho_h(b)>\rho_n(b)\) for every \(b>b_0\).
This is why \(h\) is called candidate-information-dominant and \(n\) candidate-noise-dominant.
The classification is made before differential screening.

Among completed orders under levy \(\ell<a(b)\), the informed share is
\begin{equation}
    \rho_s^E(\ell;b)
    =
    \frac{i[a(b)+d-\ell]}
         {i[a(b)+d-\ell]+n_s[a(b)-\ell]}.
    \label{eq:executed-posterior}
\end{equation}
It increases strictly with \(\ell\), because the shorter-support liquidity-motivated type is screened out faster.
This compositional statement is not, by itself, a claim about mutual information, forecast accuracy, or price informativeness.
Moreover, differential screening can reverse the cross-direction ranking after execution.

\section{SCF and the Common Retention Rule}
\label{sec:mechanism}

\subsection{A payoff-direction levy}

For an arbitrary trade \(r\) with \(x(r)\ne0\), let \(s(r)\in\{h,n\}\) denote the sign-mapped payoff direction.
For coefficients \(\tau_h,\tau_n\ge0\), define the payment
\begin{equation}
    \operatorname{Pay}_{b,\tau}(q,r)
    =
    C_b(q+r)-C_b(q)
    +\frac{\tau_{s(r)}}{2}x(r)^2.
    \label{eq:payment}
\end{equation}
Riskless trades have zero levy.
For the fixed risky lot \(z\), write
\begin{equation}
    \ell_s=\frac{\tau_s z^2}{2}.
    \label{eq:fixed-lot-levy}
\end{equation}
The quadratic form in \cref{eq:payment} applies on arbitrary trade histories; only its fixed-lot value enters the recovery theorem.

\begin{proposition}[Representation invariance and safety]
\label{prop:safety}
For every \(q,r\in\R^2\) and \(\alpha,\beta\in\R\),
\begin{equation}
\begin{aligned}
    \operatorname{Pay}_{b,\tau}(q+\beta\one,r)
    &=\operatorname{Pay}_{b,\tau}(q,r),\\
    \operatorname{Pay}_{b,\tau}(q,r+\alpha\one)
    &=\operatorname{Pay}_{b,\tau}(q,r)+\alpha.
\end{aligned}
    \label{eq:representation-identities}
\end{equation}
Thus economically equivalent implementations have identical state-contingent net profits.
Moreover, a finite history under \cref{eq:payment} admits no pathwise arbitrage, and nonnegative levies do not increase the LMSR worst-case loss.
\end{proposition}

The proposition is an implementation requirement rather than a novelty claim.
In particular, buy YES and sell NO cannot be charged differently when they represent the same risky payoff after a cash adjustment.
Because levies are nonnegative, the lower charge on direction \(h\) is fee relief rather than a rebate.

\subsection{Expected recovery revenue}

Using \cref{eq:QI,eq:QN}, expected SCF receipts are
\begin{equation}
\begin{aligned}
    \Gamma(\ell_h,\ell_n;b,m)
    &=\sum_{s\in\{h,n\}}\ell_s
       \bigl[Q_{\ind,s}+Q_{\noise,s}\bigr]\\
    &=m\sum_{s\in\{h,n\}} i d\,\ell_s\\[-2pt]
    &\quad+m\sum_{s\in\{h,n\}}
       c_s\ell_s[a(b)-\ell_s].
\end{aligned}
    \label{eq:revenue}
\end{equation}
The revenue is strictly concave under
\begin{equation}
    0\le\ell_s\le a(b).
    \label{eq:levy-cap}
\end{equation}

Retention cost is the expected number of informed orders screened out relative to zero levy.
With a uniform informed density,
\begin{equation}
    L_s(\ell_s;b)=m i\ell_s.
    \label{eq:info-loss}
\end{equation}
This completion metric does not directly measure information incorporated into prices.

The tight-loading condition is
\begin{equation}
    \bar\delta
    :=
    \frac{i(n_n-n_h)}{2(i+n_n)}
    >0,
    \qquad
    0<\delta<\bar\delta.
    \label{eq:delta}
\end{equation}
At each candidate \(b>b_0\), both mechanism classes face the same allowance
\begin{equation}
    \bar L(b)=m\delta a(b).
    \label{eq:allowance}
\end{equation}
Equivalently, the permitted loss is a constant fraction of common fee headroom.
Although pointwise fair, the rule varies with \(b\); see \cref{sec:discussion}.

The SCF recovery capacity, with \(D\) denoting its directional menu, is
\begin{equation}
\begin{aligned}
    \Gamma_D(b;m):={}&\max_{\ell_h,\ell_n}
        \ \Gamma(\ell_h,\ell_n;b,m)\\[-2pt]
    \text{s.t. }{}&i(\ell_h+\ell_n)\le\delta a(b),\\[-2pt]
                  &0\le\ell_h,\ell_n\le a(b).
\end{aligned}
    \label{eq:directional-program}
\end{equation}
The uniform benchmark chooses the best feasible common levy:
\begin{equation}
\begin{aligned}
    \Gamma_U(b;m):={}&\max_{\ell}\ \Gamma(\ell,\ell;b,m)\\[-2pt]
    \text{s.t. }{}&2i\ell\le\delta a(b),
    \quad 0\le\ell\le a(b).
\end{aligned}
    \label{eq:uniform-program}
\end{equation}
Thus both mechanisms face the same retention allowance.
Weak dominance is mechanical because the uniform class is the diagonal subset
of the SCF policy class.
The substantive question is whether strict dominance holds at every depth and expands the recoverable set.

\begin{figure*}[t]
\centering
\includegraphics[width=0.78\textwidth]{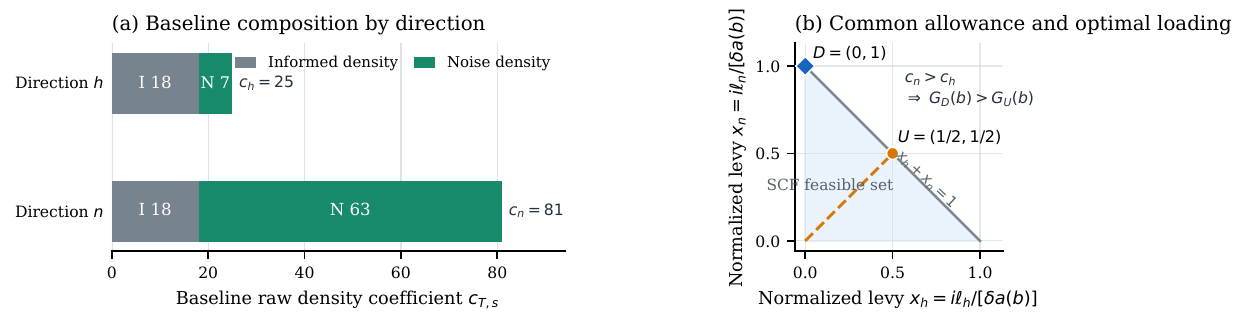}
\caption{Direction-conditioned levy allocation under a common retention budget.
(a) Informed raw density is symmetric, whereas noise is concentrated in
direction \(n\), giving \(c_h=25\) and \(c_n=81\). (b) With
\(x_s=i\ell_s/[\delta a(b)]\), SCF chooses over the shaded simplex while uniform
is restricted to its diagonal. The optima \(D=(0,1)\) and
\(U=(1/2,1/2)\) exhaust the same allowance, but directional loading yields
\(G_D(b)>G_U(b)\) for every \(b>b_0\).}
\label{fig:method-schematic}
\end{figure*}

\section{Seed-Capital Recovery Thresholds}
\label{sec:main}

We now solve \cref{eq:directional-program,eq:uniform-program} and connect the revenue comparison to prefunded seed capital.
Use the abbreviations
\begin{equation}
\begin{aligned}
    I:=2i,
    \qquad
    C:=c_h+c_n.
\end{aligned}
    \label{eq:aggregate-densities}
\end{equation}

\begin{theorem}[Informed-order-retention-constrained liquidity-set expansion]
\label{thm:main}
Suppose \cref{eq:supports,eq:flow-asymmetry,eq:delta} hold.
For every \(b>b_0\), the unique optimal SCF levies are
\begin{equation}
    \ell_h^D(b)=0,
    \qquad
    \ell_n^D(b)=\frac{\delta a(b)}{i},
    \label{eq:directional-solution}
\end{equation}
whereas the unique optimal uniform levy is
\begin{equation}
    \ell^U(b)=\frac{\delta a(b)}{2i}.
    \label{eq:uniform-solution}
\end{equation}

Let \(\Gamma_M(b;m)=mG_M(b)\) for \(M\in\{D,U\}\).
The optimized recovery functions are
\begin{equation}
    G_D(b)
    =
    \delta d\,a(b)
    +c_n a(b)^2
      \left(
        \frac{\delta}{i}
        -\frac{\delta^2}{i^2}
      \right).
    \label{eq:GD}
\end{equation}
\begin{equation}
    G_U(b)
    =
    \delta d\,a(b)
    +C a(b)^2
      \left(
        \frac{\delta}{I}
        -\frac{\delta^2}{I^2}
      \right).
    \label{eq:GU}
\end{equation}
For every \(b>b_0\),
\begin{equation}
    0<G_U(b)<G_D(b).
    \label{eq:pointwise-gap}
\end{equation}

Normalize capacity by seed capital and define
\begin{equation}
    \Psi_M(b):=\frac{G_M(b)}{b\log2},
    \qquad
    S_M:=\max_{b>b_0}\Psi_M(b).
    \label{eq:S}
\end{equation}
Both maxima are finite, positive, and attained at interior liquidity levels:
\begin{equation}
    0<S_U<S_D<\infty.
    \label{eq:S-gap}
\end{equation}

The expected seed-capital recovery sets are
\begin{equation}
    \cB_M(m)
    :=
    \left\{
        b>b_0:
        mG_M(b)\ge b\log2
    \right\}.
    \label{eq:recoverable-set}
\end{equation}
Then:
\begin{enumerate}[leftmargin=*,label=(\roman*)]
    \item if \(0<m<1/S_D\), then
    \(\cB_D(m)=\cB_U(m)=\varnothing\);
    \item if \(1/S_D\le m<1/S_U\), then
    \(\cB_D(m)\ne\varnothing\) and
    \(\cB_U(m)=\varnothing\);
    \item if \(m\ge1/S_U\), then
    \(\cB_U(m)\subsetneq\cB_D(m)\).
\end{enumerate}
\end{theorem}

\paragraph{Why SCF loads one side.}
The retention constraint in \cref{eq:directional-program} is
\begin{equation}
    \ell_h+\ell_n\le \frac{\delta a(b)}{i}.
    \label{eq:retention-total}
\end{equation}
For a fixed total levy burden, the term \(id(\ell_h+\ell_n)\) in \cref{eq:revenue} is constant.
The remaining revenue is more productive in direction \(n\), because \(c_n>c_h\).
Condition \(\delta<\bar\delta\) is strong enough that even after assigning the full allowed burden to \(n\), the marginal return there remains above the marginal return from starting to charge \(h\).
Strict concavity then gives \cref{eq:directional-solution}.
Under a common levy, the same constraint becomes \(2i\ell\le\delta a(b)\),
giving \cref{eq:uniform-solution} as the benchmark.

\paragraph{Why the gap is strict.}
The uniform solution spends half of the common screening allowance in each direction.
SCF spends all of it where the liquidity-motivated density is higher.
Both lose exactly \(m\delta a(b)\) expected informed orders, but their liquidity-motivated fee bases differ.
The strict inequality in \cref{eq:pointwise-gap} therefore arises from
\(n_n>n_h\), a binding common constraint, and strict concavity, not merely from
nesting the uniform menu in the SCF policy class.

\paragraph{From pointwise revenue to launchability.}
The seed-capital-normalized ratio \(\Psi_M\) tends to zero at both ends of its domain.
As \(b\downarrow b_0\), fee headroom and revenue vanish.
As \(b\to\infty\), \(a(b)\to w_N\), so \(G_M(b)\) remains bounded while the seed \(b\log2\) diverges.
Continuity yields an interior maximum.
At a maximizer of \(\Psi_U\), pointwise strictness implies \(\Psi_D>\Psi_U\), and hence \(S_D>S_U\).
The three threshold cases then follow from
\begin{equation}
\begin{aligned}
    b\in\cB_M(m)
    \quad\Longleftrightarrow\quad
    m\Psi_M(b)\ge1.
\end{aligned}
    \label{eq:recoverability-equivalence}
\end{equation}

\paragraph{Robustness to a liquidity-invariant allowance.}
The pointwise allowance in \cref{eq:allowance} varies with \(b\).
Appendix~\ref{app:fixed-allowance} instead fixes the allowed informed-order loss
per unit of arrival scale at \(\Lambda\) on a domain
\(b\ge\underline b>b_0\). If
\(0<\Lambda<\bar\delta a(\underline b)\), the SCF solution still
loads the entire burden onto \(n\), strictly dominates the optimal uniform
levy pointwise, and preserves all three arrival-threshold conclusions.
The positive liquidity floor is necessary because fee headroom vanishes as
\(b\downarrow b_0\).

\begin{proposition}[Symmetric benchmark]
\label{prop:symmetry}
Maintain the linear-surplus opening-cohort model, but replace
\cref{eq:flow-asymmetry,eq:delta} with \(i_h=i_n=:i>0\),
\(n_h=n_n=:n>0\), and a direction-symmetric retention allowance. Specifically,
for some \(\bar L(b)\ge0\), let SCF face
\(mi(\ell_h+\ell_n)\le\bar L(b)\) with identical caps in both directions, and
let the uniform program face \(2mi\ell\le\bar L(b)\). Then the SCF objective
and feasible set are invariant to exchanging \(h\) and \(n\). The unique SCF
optimizer is diagonal, and the optimized SCF and uniform recovery capacities
coincide.
\end{proposition}

\begin{remark}[Candidate versus executed composition]
Theorem~\ref{thm:main} classifies direction \(n\) using the zero-levy candidate posterior in \cref{eq:candidate-posterior}.
Because \(\rho_s^E(\ell;b)\) rises with the levy, the higher charge can make completed orders in direction \(n\) more informed on average and can even reverse the original ranking.
The theorem does not require the ex-post ranking to remain fixed.
\end{remark}

\begin{remark}[What is optimal]
The word \emph{optimal} always means optimal within the capped class of nonnegative per-lot levies in \cref{eq:directional-program,eq:uniform-program}, for the opening-cohort revenue objective and the stated retention allowance.
It does not mean globally optimal market design.
\end{remark}

\section{Experiments}
\label{sec:experiments}

The theorem is analytical; the experiments are synthetic comparative statics
and falsification checks \newtext{\citep{liu2026finsimsurvey}}, not empirical evidence. For each parameter vector we
globally maximize
\(\Psi_M(b)=G_M(b)/(b\log2)\) over \(b>b_0\) using a 512-point
adaptive log grid and golden-section refinement. The primary outcome is the
proportional reduction in the minimum arrival scale,
\begin{equation}
    R:=1-\frac{m_D^\star}{m_U^\star}
      =1-\frac{S_U}{S_D}.
    \label{eq:threshold-reduction}
\end{equation}
Positive \(R\) favors SCF. No main-model maximum is on a numerical boundary,
and the baseline agrees with a separate 20,001-point grid within
\(3.5\times10^{-9}\) in normalized capacity. Parameter-draw percentiles below
describe the chosen synthetic design measure; they are not confidence
intervals or estimates of a market population.

\begin{figure*}[!t]
\centering
\includegraphics[width=0.8\textwidth]{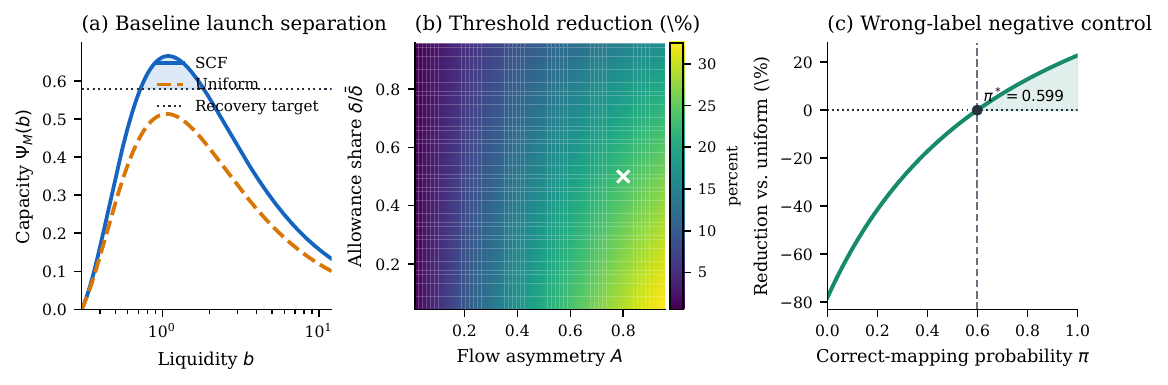}
\caption{Synthetic audit. (a) Baseline seed-normalized recovery curves; shading
marks liquidity levels recoverable only under SCF at the displayed arrival
scale. (b) Threshold reduction over flow asymmetry and retention tightness; the
cross marks the baseline. (c) Exact wrong-label negative control.}
\label{fig:experiment-summary}
\end{figure*}

\paragraph{Baseline.}
We set \(z=1\), \((w_I,w_N)=(0.45,0.30)\),
\((i_h,i_n)=(18,18)\), \((n_h,n_n)=(7,63)\), and
\(\delta=\bar\delta/2=3.111111\). This gives
\begin{equation}
\begin{aligned}
 b_0&=0.304759, & S_D^{-1}&=1.503569,\\[-2pt]
 R&=22.84\%, & S_U^{-1}&=1.948531.
\end{aligned}
\label{eq:baseline-thresholds}
\end{equation}
At \(m=1.726050\), SCF crosses the recovery target while the uniform
policy does not (\cref{fig:experiment-summary}a). One unit of \(m\) contains
37.2 expected raw candidate orders under this calibration, so the two
thresholds correspond to 55.93 and 72.49 candidates before participation
screening; this conversion does not turn \(m\) into observed traffic.

\paragraph{Independent program and boundary audits.}
The large sweeps inside the theorem domain reuse its closed-form capacity and
therefore cannot independently validate the optimizer. We separately solve the
raw two-levy programs in \cref{eq:directional-program,eq:uniform-program},
without calling the closed-form capacity routine, over 20 designs spanning four
calibrations and five fee-headroom levels. A complete 1,001-point feasible-edge
search with local refinement matches the closed-form levies within
\(3.34\times10^{-16}\), capacities within \(2.22\times10^{-16}\), and the
common loss budget within \(2.22\times10^{-16}\).

We then keep the baseline normalized loss budget but include signed noise-flow
asymmetry \(A\in\{-0.8,0,0.01,0.2,0.8\}\). At \(A=0\), the numerical optimum
is diagonal and \(R=4.44\times10^{-16}\); swapping \(A=0.8\) to \(-0.8\)
leaves \(R=22.84\%\) unchanged and swaps the charged side. At \(A=0.01\), the
gain is only \(0.0073\%\). Finally, holding \(A=0.8\) fixed and moving the
allowance share from \(q=0.99\) to \(q=1.25\) changes the optimum from
one-sided to split loading. The gain remains nonnegative because the optimized
directional menu contains the uniform menu, but the one-sided prescription is
not extended beyond its stated tight-loading condition.

\begin{table}[t]
\centering
\footnotesize
\setlength{\tabcolsep}{2.0pt}
\begin{tabular}{lll}
\toprule
Audit & Setting & Result \\
\midrule
Raw program & 20 designs & \(\max|\Delta G|=2.22\!\times\!10^{-16}\) \\
Symmetry & \(A=0\) & \(R\doteq0\) \\
Side swap & \(A=\pm0.8\) & same \(R\), levies swap \\
Loading & \(q:.99\to1.25\) & \(n\)-only \(\to\) split \\
Recovery sets & low/mid/high \(m\) & \((0,0),(1,0),(1,1)\) \\
\bottomrule
\end{tabular}
\caption{Audits that can contradict the closed-form implementation or its
boundary interpretation. Recovery-set entries count nonempty SCF and uniform
sets, respectively. Full rows and residuals are in the supplement.}
\label{tab:independent-audits}
\end{table}

\paragraph{Recoverable sets and structured sweep.}
Root refinement directly recovers the three cases of \cref{thm:main}. At
\(m=1.353212\), both sets are empty. At \(m=1.726050\), the SCF set is
\([0.724392,1.828707]\) and the uniform set is empty. At \(m=2.143384\), the
SCF and uniform sets are \([0.591848,2.664094]\) and
\([0.753474,1.625315]\), respectively; the latter is strictly contained in
the former. Maximum endpoint residual is \(1.34\times10^{-15}\).

For comparative statics, define flow asymmetry
\(A=(n_n-n_h)/(n_n+n_h)\) and retention tightness
\(q=\delta/\bar\delta\). A \(64\times61\) grid over
\(A\in[0.02,0.95]\) and \(q\in[0.05,0.95]\), holding mean noise density
fixed, produces a smooth phase diagram (\cref{fig:experiment-summary}b).
All 3,904 cells have \(R>0\), as the theorem requires, while the magnitude
shrinks toward symmetry.

\begin{figure}[t]
\centering
\includegraphics[width=0.7\columnwidth]{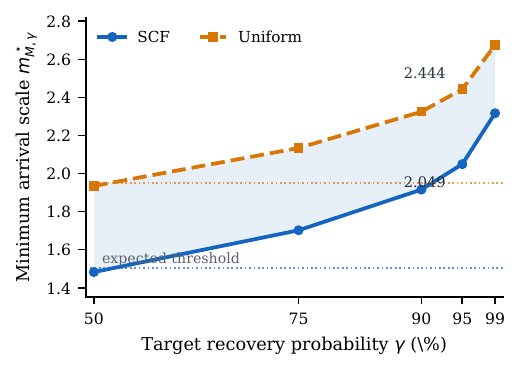}
\caption{Exact finite-cohort thresholds under independent Poisson thinning.
Each mechanism chooses \(b\) before the opening cohort realizes. Dotted lines
mark the corresponding expected-receipt thresholds.}
\label{fig:poisson-chance}
\end{figure}

\paragraph{Finite-cohort recovery probability.}
Expected intensities do not specify realized order counts. Let the four
type--direction cells be independent Poisson point processes over the recovery
window. Thinning by participation and locking the levies before orders realize
gives \(V_M=\ell_M\operatorname{Pois}(m\lambda_M(b))\), with
\(\E[V_M]=mG_M(b)\). For each target probability \(\gamma\), each mechanism
chooses \(b\) ex ante to minimize the required \(m\). At \(\gamma=0.95\), the
thresholds are 2.0494 for SCF and 2.4436 for uniform, a 16.13\% reduction
(\cref{fig:poisson-chance}). The reduction falls from 23.40\% at
\(\gamma=0.50\) to 13.40\% at \(\gamma=0.99\).

Independent Poisson flow is thin-tailed. As a stress test, multiply all four
cell intensities in a window by a common
\(U\sim\operatorname{Gamma}(k,k)\), preserving their means while inducing
clustering. At the 95\% target, \(k=10\) raises the SCF and uniform thresholds
to 2.9706 and 3.7256; \(k=2\) raises them to 8.5794 and 11.0436. Thus common
shocks sharply worsen absolute launch traffic, while the relative reductions
remain 20.27\% and 22.31\%. This check covers a shared multiplicative shock,
not arbitrary directional dependence.

\paragraph{Implementation coverage and distributional stress.}
A 324-case factorial and 10,000 fixed-seed parameter draws vary
\(w_N/z\), the normalized surplus gap, mean noise-to-informed density, \(A\),
and \(q\) over the theorem's admissible interior. All 10,324 cases retain the
strict ordering, with \(R\) ranging from \(0.016\%\) to \(43.04\%\). Because
strictness is implied by the theorem and the draw distribution is artificial,
this is implementation coverage rather than independent or statistical
evidence.

To move outside the uniform-surplus assumption, we replace each type's
normalized surplus density by Beta\((1,3)\), Beta\((1,1)\), Beta\((2,2)\),
or Beta\((3,1)\), preserving cell mass, informed symmetry, and the common
absolute loss allowance. Tail-certified multistart reoptimization in all 16
pairs retains \(S_D>S_U\), with \(R=0.030\)--38.35\%. The one-sided pattern
survives in only 10 cases; high-surplus informed flow with center-concentrated
noise is the near-counterexample. This finite search is not a theorem
extension.

\paragraph{Wrong-label negative control.}
The theorem assumes the noise-dominant direction is known. Let \(\pi\) be the
probability that a recovery window uses the correct direction mapping and let
\(r=\delta/i\). The exact baseline break-even is
\begin{equation}
 \pi^*=\frac{C(r/2-r^2/4)-c_h(r-r^2)}
 {(c_n-c_h)(r-r^2)}=0.598881.
 \label{eq:label-break-even}
\end{equation}
SCF beats uniform only when \(\pi>\pi^*\)
(\cref{fig:experiment-summary}c). Complete reversal raises the baseline
arrival threshold by 78.17\% relative to uniform. The value 59.89\% is not
universal: across the 10,324 admissible designs, \(\pi^*\) has median 61.17\%,
5th--95th percentiles 51.36--72.77\%, and range 50.34--74.47\%. Here \(\pi\)
is a window-level mapping probability, not per-order classifier accuracy.

\paragraph{Fixed allowance and fragmentation.}
For four positive minimum depths and three tightness levels, the
liquidity-invariant allowance of \cref{prop:fixed-allowance} preserves a
17.78\%--27.08\% threshold reduction. In contrast, splitting a quadratically
charged lot into \(k\) equal pieces leaves only \(1/k\) of the levy, supporting
the indivisible-lot domain but not Sybil robustness.

\section{Discussion}
\label{sec:discussion}

\paragraph{What kind of liquidity is added?}
The mechanism does not manufacture reserves after a market has opened.
The sponsor supplies the full LMSR seed before trading.
Conditional on a positive potential opening cohort, the fee allocation can make
more candidate values of \(b\) recoverable in expectation.
This is a financing or launchability result: if \(m=0\), levy revenue is zero
under every policy and no positive \(b\) is recoverable.

\paragraph{Pointwise fairness and cross-\(b\) normalization.}
At each candidate \(b\), both mechanisms face the same allowance
\(\bar L(b)=m\delta a(b)\), but this allowed number of screened informed orders
varies with depth. Proposition~\ref{prop:fixed-allowance} shows that the launch
separation survives a fixed per-arrival allowance on a domain bounded away from
the zero-headroom boundary. Neither result establishes robustness to every
information-quality or retention objective.

\section{Conclusion}
\label{sec:conclusion}

SCF does not solve a literal no-trade cold start.
It can nevertheless change the financing boundary of a prefunded prediction
market when potential order flow is positive and direction predicts order
composition. In the binary LMSR opening-cohort model, a tight common
informed-order-retention rule makes it optimal to exempt the
candidate-information-dominant direction and concentrate the permitted charge
on the candidate-noise-dominant direction. Relative to the best uniform levy
under the same rule, SCF has strictly higher expected recovery capacity
at every admissible liquidity level and lowers the launch threshold.

The contribution links a common screening constraint to an exact capital
threshold; fees, directional pricing, and noise-flow revenue are not new by
themselves. A natural extension is a sequential model with an
information-quality objective.

\section{Limitations}
\label{sec:limitations}

The analysis is limited to a fixed opening cohort in a binary LMSR. The
retention constraint counts informed orders rather than signal quality and does
not establish sequential equilibrium, price accuracy, or endogenous direction
choice. Ex-ante direction labels may be misspecified, and no empirical
classifier is evaluated. Recovery is expected rather than pathwise; the
Poisson and shared-shock checks omit directional clustering and serial
dependence. The quadratic levy is vulnerable to trade splitting. Other
settings require separate analysis.

\begingroup
\parfillskip=0pt plus .75\linewidth
\pretocmd{\bibitem}{\looseness=-1}{}{}
\bibliography{references}
\endgroup

\appendix
\section{Proofs}
\label{app:proofs}

\subsection{Proof of the LMSR-geometry lemma}

Put \(x=z/(2b)>0\).
Differentiating \cref{eq:D} gives
\[
    D'(b)=\log\cosh x-x\tanh x.
\]
As a function of \(x\), the right-hand side has derivative
\[
    -x\operatorname{sech}^2x<0
\]
and limit zero at \(x=0\).
It is therefore negative for every \(x>0\), so \(D\) is strictly decreasing in \(b\).
Using \(\log\cosh x=x-\log2+o(1)\) as \(x\to\infty\) gives
\[
    \lim_{b\downarrow0}D(b)=\frac z2.
\]
Using \(\log\cosh x=x^2/2+o(x^2)\) as \(x\downarrow0\) gives
\[
    D(b)=\frac{z^2}{8b}+o(b^{-1})
\]
and hence \(D(b)\to0\) as \(b\to\infty\), completing the uniqueness argument.
\qed

\subsection{Proof of Proposition~\ref{prop:safety}}

Translation invariance of \(C_b\) gives
\[
    C_b(q+\beta\one+r)-C_b(q+\beta\one)
    =
    C_b(q+r)-C_b(q).
\]
Because \(x(r+\alpha\one)=x(r)\), the levy is unchanged by either a state translation or a riskless overlay.
Translating the traded claim gives
\[
    C_b(q+r+\alpha\one)-C_b(q)
    =
    C_b(q+r)-C_b(q)+\alpha,
\]
which proves the two payment identities.
The terminal payoff of the overlay also rises by \(\alpha\) in each outcome, so state-contingent net profit is unchanged.

For pathwise no arbitrage, consider a finite history \(q_{t+1}=q_t+r_t\).
Exact cost differences telescope:
\begin{align*}
    \sum_t\operatorname{Pay}_{b,\tau}(q_t,r_t)
    &=
    C_b(q_T)-C_b(q_0)\\
    &\quad+\sum_{t:x(r_t)\ne0}\frac{\tau_{s(r_t)}}2x(r_t)^2\\
    &\ge C_b(q_T)-C_b(q_0).
\end{align*}
Define the net position, total levy, and minimum component by
\[
    v=q_T-q_0,
    \qquad
    v_{\min}=\min\{v_Y,v_N\}.
\]
\[
\begin{aligned}
    L
    &:=\sum_{t:x(r_t)\ne0}
       \frac{\tau_{s(r_t)}}2x(r_t)^2
       \ge0.
\end{aligned}
\]
Let
\[
\begin{aligned}
    p_j(q_0)
    &:=\frac{e^{q_{0,j}/b}}
      {e^{q_{0,Y}/b}+e^{q_{0,N}/b}}\\
    &\in(0,1),
      \qquad j\in\{Y,N\}.
\end{aligned}
\]
If \(v_Y\ne v_N\), then
\[
\begin{aligned}
    C_b(q_0+v)-C_b(q_0)
    &=b\log\!\Bigl(
        p_Y(q_0)e^{v_Y/b}\\[-2pt]
    &\qquad\quad
        +p_N(q_0)e^{v_N/b}
      \Bigr)\\
    &>v_{\min},
\end{aligned}
\]
where strictness follows because both LMSR state prices are positive.
The aggregate net payoff in an outcome attaining \(v_{\min}\) is therefore
strictly negative after including \(L\). If instead
\(v_Y=v_N=v_{\min}\), translation invariance gives
\[
    C_b(q_0+v)-C_b(q_0)=v_{\min},
\]
so aggregate net payoff equals \(-L\le0\) in both outcomes. Thus no finite
history has nonnegative net payoff in every outcome and strictly positive net
payoff in at least one outcome.

Finally, starting from \(q^0=0\), the worst-case loss in outcome \(Y\) is
\begin{align*}
    &\sup_q\{q_Y-[C_b(q)-C_b(0)]\}\\
    &=
    b\log2+\sup_q\{q_Y-C_b(q)\}\\
    &=
    b\log2,
\end{align*}
where the last supremum is zero and is approached as \(q_Y-q_N\to\infty\).
The same calculation holds in outcome \(N\).
Every nonnegative levy adds receipts and hence cannot increase this loss.
\qed

\subsection{Proof of the main theorem}

Fix \(b>b_0\) and abbreviate \(a=a(b)\).
Divide expected revenue by \(m\), and write \(x=\ell_h\), \(y=\ell_n\).
The SCF objective is
\begin{equation}
    F(x,y)
    =
    id(x+y)
    +c_hx(a-x)
    +c_ny(a-y).
    \label{eq:appendix-F}
\end{equation}
Its Hessian is
\[
    \nabla^2F
    =
    -2\begin{pmatrix}c_h&0\\0&c_n\end{pmatrix},
\]
so \(F\) is strictly concave.
The retention constraint is
\begin{equation}
    x+y\le E,
    \qquad
    E:=\frac{\delta a}{i}.
    \label{eq:E}
\end{equation}
Condition \cref{eq:delta} implies
\[
    \frac{\delta}{i}
    <
    \frac{n_n-n_h}{2(i+n_n)}
    <\frac12,
\]
so \(0<E<a/2\).
Both partial derivatives of \(F\) are positive throughout the feasible triangle:
\begin{align*}
    \frac{\partial F}{\partial x}&=id+c_h(a-2x)>0,\\
    \frac{\partial F}{\partial y}&=id+c_n(a-2y)>0.
\end{align*}
The constraint in \cref{eq:E} must therefore bind.

Substitute \(y=E-x\) and differentiate:
\[
    \frac{d}{dx}F(x,E-x)
    =
    a(c_h-c_n)+2c_nE-2(c_h+c_n)x.
\]
At \(x=0\), this derivative is negative exactly when
\[
    \delta
    <
    \frac{i(c_n-c_h)}{2c_n}
    =
    \frac{i(n_n-n_h)}{2(i+n_n)}
    =\bar\delta.
\]
It decreases further with \(x\), so strict concavity yields the unique SCF
maximum \(x=0\), \(y=E\) in \cref{eq:directional-solution}.

Under a uniform levy \(\ell\), revenue per unit of arrival scale is
\[
    F_U(\ell)=2id\,\ell+C\ell(a-\ell),
\]
and feasibility requires \(0\le\ell\le E/2\).
Because \(E/2<a/4\),
\[
    F_U'(\ell)=2id+C(a-2\ell)>0
\]
over the feasible interval.
The unique uniform maximum is \(\ell=E/2\), proving \cref{eq:uniform-solution}.

Substituting the two optimizers into \cref{eq:revenue} gives
\cref{eq:GD,eq:GU}.
Both are positive because \(a,d,\delta>0\) and
\[
    0<\frac{\delta}{i}<\frac12,
    \qquad
    0<\frac{\delta}{I}<\frac14.
\]
The uniform optimizer corresponds to a feasible diagonal point in the SCF program.
The SCF optimizer is unique and non-diagonal.
Strict concavity therefore implies
\[
    G_D(b)>G_U(b).
\]
This proves \cref{eq:pointwise-gap}.

We next establish the seed-capital-normalized claims.
Both \(G_D\) and \(G_U\) are continuous on \((b_0,\infty)\).
By \cref{lem:lmsr-geometry}, \(a(b)\downarrow0\) as \(b\downarrow b_0\), so
\[
    \lim_{b\downarrow b_0}\Psi_M(b)=0.
\]
As \(b\to\infty\), \(a(b)\to w_N\).
Each \(G_M(b)\) therefore approaches a finite constant, whereas \(b\log2\to\infty\), so
\[
    \lim_{b\to\infty}\Psi_M(b)=0.
\]
Because \(\Psi_M\) is continuous and positive in the interior, it attains a finite positive maximum at an interior point.
Let \(b_U^\star\) maximize \(\Psi_U\).
Pointwise strictness gives
\[
    S_D
    \ge \Psi_D(b_U^\star)
    >
    \Psi_U(b_U^\star)
    =S_U,
\]
proving \cref{eq:S-gap}.

The recoverability condition is
\[
    b\in\cB_M(m)
    \quad\Longleftrightarrow\quad
    m\Psi_M(b)\ge1.
\]
Below \(1/S_D\), neither mechanism reaches the target.
If \(1/S_D\le m<1/S_U\), attainment of \(S_D\) makes the SCF recovery set
nonempty, whereas the uniform maximum remains below the target.

Suppose \(m\ge1/S_U\) and set \(q=1/m\).
Then \(0<q\le S_U<S_D\).
Pointwise strictness immediately gives
\(\cB_U(m)\subseteq\cB_D(m)\).
The uniform superlevel set
\[
    A_U=\{b>b_0:\Psi_U(b)\ge q\}
\]
is nonempty and compact because \(\Psi_U\) is continuous and tends to zero at both endpoints.
Let \(\bar b=\sup A_U\).
Continuity gives \(\Psi_U(\bar b)=q\), while strictness gives
\(\Psi_D(\bar b)>q\).
Immediately to the right of \(\bar b\), continuity leaves
\(\Psi_D>q\), whereas the definition of the right endpoint gives
\(\Psi_U<q\).
Thus some point lies in
\(\cB_D(m)\setminus\cB_U(m)\), so the inclusion is strict.
\qed

\section{Fixed Loss-Allowance Robustness}
\label{app:fixed-allowance}

\begin{proposition}[Liquidity-invariant per-arrival loss allowance]
\label{prop:fixed-allowance}
Maintain the assumptions of \cref{thm:main}. Fix a minimum candidate liquidity
\(\underline b>b_0\) and restrict attention to
\(b\in[\underline b,\infty)\). Replace \cref{eq:allowance} with
\begin{equation}
    \bar L_{\Lambda}(m)=m\Lambda,
    \qquad
    0<\Lambda<\bar\delta\,a(\underline b),
    \label{eq:fixed-allowance}
\end{equation}
which is constant across candidate liquidity levels per unit of arrival scale.
Here \(\Lambda\) is an informed-order-loss allowance per unit of arrival scale,
not an aggregate allowance independent of \(m\) or a fixed retained fraction.
For every \(b\ge\underline b\), the unique optima are
\begin{equation}
\begin{aligned}
    \ell_h^{D,\Lambda}(b)&=0,
    &\ell_n^{D,\Lambda}(b)&=\frac{\Lambda}{i},\\[-2pt]
    \ell^{U,\Lambda}(b)&=\frac{\Lambda}{2i}.&&
\end{aligned}
    \label{eq:fixed-allowance-solutions}
\end{equation}
Their per-unit-arrival recovery functions are
\begin{align}
    G_D^{\Lambda}(b)
    &=d\Lambda+c_n\frac{\Lambda}{i}
       \left(a(b)-\frac{\Lambda}{i}\right),
    \label{eq:fixed-GD}\\
    G_U^{\Lambda}(b)
    &=d\Lambda+C\frac{\Lambda}{2i}
       \left(a(b)-\frac{\Lambda}{2i}\right).
    \label{eq:fixed-GU}
\end{align}
They satisfy \(0<G_U^{\Lambda}(b)<G_D^{\Lambda}(b)\) for every
\(b\ge\underline b\). For each mechanism, set
\begin{align*}
    S_M^{\Lambda}&:=\max_{b\ge\underline b}
       \frac{G_M^{\Lambda}(b)}{b\log2},\\
    \cB_M^{\Lambda}(m)&:=
    \left\{b\ge\underline b:
    mG_M^{\Lambda}(b)\ge b\log2\right\}.
\end{align*}
Then \(0<S_U^{\Lambda}<S_D^{\Lambda}<\infty\), and all three
arrival-threshold and strict-set conclusions of \cref{thm:main} hold after
replacing \((S_M,\cB_M)\) by
\((S_M^{\Lambda},\cB_M^{\Lambda})\).
\end{proposition}

\subsection{Proof of Proposition~\ref{prop:fixed-allowance}}

By \cref{lem:lmsr-geometry}, \(D(b)\) is strictly decreasing, so
\(a(b)=w_N-D(b)\) is strictly increasing.
Hence, for every \(b\ge\underline b\),
\[
    \Lambda<\bar\delta a(\underline b)
    \le \bar\delta a(b).
\]
Fix such a \(b\), abbreviate \(a=a(b)\), and set
\[
    E_{\Lambda}:=\frac{\Lambda}{i}.
\]
The SCF retention constraint is
\(x+y\le E_{\Lambda}\).
As in the proof of \cref{thm:main}, the bound on \(\Lambda\) implies
\(0<E_{\Lambda}<a/2\), both partial derivatives of
\cref{eq:appendix-F} are positive on the feasible triangle, and the
retention constraint binds.

After substituting \(y=E_{\Lambda}-x\), the derivative at \(x=0\) is
\[
    a(c_h-c_n)+2c_nE_{\Lambda}<0,
\]
where the inequality is equivalent to
\(\Lambda<\bar\delta a\).
The derivative decreases with \(x\), so strict concavity gives the unique
SCF optimum \((x,y)=(0,E_{\Lambda})\).
The uniform objective is strictly increasing on
\([0,E_{\Lambda}/2]\), giving the unique common levy
\(E_{\Lambda}/2\).
This proves \cref{eq:fixed-allowance-solutions}.
Substitution into \cref{eq:revenue} yields
\cref{eq:fixed-GD,eq:fixed-GU}.

The uniform optimum is a feasible diagonal point of the SCF
problem, whereas the unique SCF optimum is non-diagonal.
Strict concavity therefore gives
\(0<G_U^{\Lambda}(b)<G_D^{\Lambda}(b)\) pointwise on the restricted
domain.
The two seed-capital-normalized ratios are continuous and positive on
\([\underline b,\infty)\), and they tend to zero as \(b\to\infty\)
because their numerators remain bounded while \(b\log2\) diverges.
Consequently both positive finite maxima are attained, and evaluating the
SCF ratio at a uniform maximizer gives
\(S_D^{\Lambda}>S_U^{\Lambda}\).
Attainment of both maxima gives the first two arrival-threshold cases,
including the equality at \(1/S_D^{\Lambda}\).
For the strict-inclusion case, let
\(q=1/m\le S_U^{\Lambda}\) and define
\[
    A_U^{\Lambda}(q)
    =
    \left\{
        b\ge\underline b:
        \frac{G_U^{\Lambda}(b)}{b\log2}\ge q
    \right\}.
\]
Let \(\bar b\) be the finite right endpoint of this nonempty compact set.
Continuity gives
\(G_U^{\Lambda}(\bar b)/(\bar b\log2)=q\), whereas pointwise
strictness gives \(G_D^{\Lambda}(\bar b)/(\bar b\log2)>q\).
Immediately to the right, the SCF ratio remains above \(q\) while
the uniform ratio is below it.
Hence
\(\cB_U^{\Lambda}(m)\subsetneq\cB_D^{\Lambda}(m)\).
\qed

\subsection{Proof of Proposition~\ref{prop:symmetry}}

Under the symmetric benchmark, exchanging \(h\) and \(n\) leaves the SCF
objective and feasible set unchanged.
Strict concavity gives a unique optimizer.
The exchanged optimizer must therefore equal the original one, implying
\(\ell_h^D=\ell_n^D\).
This diagonal solution is feasible for the uniform problem with the same objective value, so the two capacities coincide.
\qed

\section{Additional Calculations and Extensions}
\label{app:extensions}

\subsection{Executed-order posterior}

Use the shorthand
\[
    H_s(\ell;b):=i[a(b)+d-\ell]+n_s[a(b)-\ell].
\]
The derivative of \cref{eq:executed-posterior} is
\[
    \frac{\partial \rho_s^E(\ell;b)}{\partial\ell}
    =\frac{i n_s d}{H_s(\ell;b)^2}>0.
\]
At zero levy, the difference \(\rho_h(b)-\rho_n(b)\) is positive because \(n_n>n_h\).
No analogous cross-direction ranking is asserted after applying different levies.

\subsection{Mechanism-independent screening}

The allocation part of \cref{thm:main} does not depend on the logarithmic form by itself.
Consider a base price-plus-fee mechanism \(B\) with:
\begin{enumerate}[leftmargin=*,label=(\alph*)]
    \item a common opening-state participation cost \(D_B(b)\) for both fixed-lot payoff directions;
    \item a prefunded capital requirement \(K_B(b)>0\);
    \item a nonnegative recovery levy that does not increase that capital requirement; and
    \item the same fixed raw type--direction primitives as in \cref{sec:model}.
\end{enumerate}
On the domain where
\[
    a_B(b):=w_N-D_B(b)>0,
\]
the constrained levy programs yield
\[
    \ell_h^D=0,
    \qquad
    \ell_n^D=\frac{\delta a_B(b)}{i},
    \qquad
    \ell^U=\frac{\delta a_B(b)}{2i},
\]
and the same strict pointwise ordering after replacing \(a\) by \(a_B\) in
\cref{eq:GD,eq:GU}.
If, in addition, the ratios
\[
    \Psi_M^B(b)=\frac{G_M^B(b)}{K_B(b)}
\]
are continuous, vanish at the relevant domain endpoints, and attain positive interior maxima, the threshold and strict-set arguments in \cref{thm:main} carry over verbatim.

This reduction explains how the screening logic may be examined with CAPM- or SQPM-style base fees: their base fee enters \(D_B\), while their capital guarantee enters \(K_B\).
It is not a claim that those mechanisms automatically satisfy every condition above.
Their participation, safety, information-incorporation, and fragmentation properties must be verified separately.

\subsection{Why unrestricted splitting defeats a quadratic recovery levy}

For a risky trade with \(x(r)=z\), the quadratic levy is \(\tau_s z^2/2\).
If the trader can split it into \(k\) consecutive pieces of risky size \(z/k\) without any aggregation rule, total additional levy is
\[
    k\frac{\tau_s}{2}\left(\frac zk\right)^2
    =
    \frac{\tau_s z^2}{2k}
    \longrightarrow0.
\]
Thus a positive anonymous quadratic recovery levy cannot simultaneously preserve its stated fixed-lot revenue and be robust to arbitrary fragmentation.

\section{Reproducibility}
\label{app:reproducibility}

The anonymous supplementary material contains the reference implementation.
The deterministic audit can be reproduced with:
\begin{verbatim}
python3 -m unittest discover \
  -s tests -v
python3 run_sanity.py \
  --output-dir results
python3 run_experiments.py \
  --output-dir results
\end{verbatim}
The frozen result files generate the paper figures:
\begin{verbatim}
MPLBACKEND=Agg python3 make_figures.py
\end{verbatim}

\subsection{Frozen experiment designs}

The independent program audit reconstructs demand and revenue directly from
\cref{eq:QI,eq:QN,eq:revenue}. For each fixed \(b\), it enumerates the complete
feasible edge of the directional loss-budget set with 1,001 points, locally
refines the best basin, and separately maximizes the uniform objective over its
exact feasible interval. It does not call the theorem's closed-form capacity
function. The 20 cases cross four calibrations (baseline, near symmetry, low
loading, and \(q=0.99\)) with five headroom levels. The maximum absolute levy,
capacity, and budget-residual errors are \(3.34\times10^{-16}\),
\(2.22\times10^{-16}\), and \(2.22\times10^{-16}\), respectively.

The boundary controls use the same raw-program solver. Signed asymmetry holds
mean noise density and the baseline normalized loss budget fixed, including
the symmetry null and a label permutation. The loading audit holds baseline
asymmetry fixed while moving the budget across the theorem's tight-loading
boundary. Selected results appear in \cref{tab:boundary-controls}; every
reported levy is normalized by \(a(b_D^\star)\).

\begin{table}[t]
\centering
\footnotesize
\setlength{\tabcolsep}{2.8pt}
\begin{tabular}{llrrrl}
\toprule
Audit & Level & \(R\) (\%) & \(\ell_h/a\) & \(\ell_n/a\) & Allocation \\
\midrule
Signed \(A\) & \(-0.80\) & 22.836 & .17284 & .00000 & \(h\)-only \\
Signed \(A\) & \(0\) & 0.000 & .08642 & .08642 & split \\
Signed \(A\) & \(0.01\) & 0.007 & .08369 & .08915 & split \\
Signed \(A\) & \(0.20\) & 2.829 & .03180 & .14104 & split \\
Signed \(A\) & \(0.80\) & 22.836 & .00000 & .17284 & \(n\)-only \\
Loading \(q\) & \(0.99\) & 13.802 & .00000 & .34222 & \(n\)-only \\
Loading \(q\) & \(1.25\) & 8.964 & .06604 & .36606 & split \\
Loading \(q\) & \(2.00\) & 2.023 & .26415 & .42721 & split \\
\bottomrule
\end{tabular}
\caption{Raw-program assumption-boundary controls. A fully optimized
directional menu weakly dominates uniform even outside the one-sided-loading
condition because it contains the uniform menu.}
\label{tab:boundary-controls}
\end{table}

For the recoverable-set audit, an 8,193-point adaptive log grid locates every
sign change of \(m\Psi_M(b)-1\), after which each endpoint is refined by
bisection. \Cref{tab:recovery-sets} reports all three theorem regimes. A dash
means the set is empty.

\begin{table}[t]
\centering
\footnotesize
\setlength{\tabcolsep}{3.5pt}
\begin{tabular}{llrr}
\toprule
\(m\) & Policy & Lower \(b\) & Upper \(b\) \\
\midrule
1.353212 & SCF & -- & -- \\
1.353212 & Uniform & -- & -- \\
1.726050 & SCF & 0.724392 & 1.828707 \\
1.726050 & Uniform & -- & -- \\
2.143384 & SCF & 0.591848 & 2.664094 \\
2.143384 & Uniform & 0.753474 & 1.625315 \\
\bottomrule
\end{tabular}
\caption{Direct numerical recovery-set audit. The maximum root residual is
\(1.34\times10^{-15}\).}
\label{tab:recovery-sets}
\end{table}

The expected-receipt suite uses the 512-point adaptive log grid and
golden-section refinement stated in \cref{sec:experiments}. The structured grid
has 64 asymmetry values and 61 allowance shares. The factorial and random
designs contain 324 and 10,000 cases, respectively; the latter uses seed
20260812. All parameter vectors and unrounded outputs are stored as CSV files.

For the finite-cohort experiment, write
\(V_M=\ell_M\operatorname{Pois}(m\lambda_M(b))\). Under the theorem's levies,
the charged-order intensities per unit \(m\) are
\begin{equation}
\begin{aligned}
 \lambda_D&=i(a+d-\ell_D)+n_n(a-\ell_D),\\[-2pt]
 \lambda_U&=2i(a+d-\ell_U)+(n_h+n_n)(a-\ell_U),
\end{aligned}
\label{eq:poisson-intensities}
\end{equation}
where \(\ell_D=\delta a/i\) and \(\ell_U=\delta a/(2i)\). Define
\(k_M(b):=\lceil K(b)/\ell_M(b)\rceil\). Independent Poisson thinning of the
opening state gives
\begin{equation}
 \Pr[V_M\ge K(b)]
 =\Pr\!\left[
   \operatorname{Pois}(m\lambda_M(b))\ge k_M(b)
 \right].
 \label{eq:poisson-tail-appendix}
\end{equation}
For each \(\gamma\), define
\begin{equation}
 m^\star_{M,\gamma}
 =\inf_{b>b_0}\frac{\mu_\gamma(k_M(b))}{\lambda_M(b)},
 \label{eq:chance-threshold}
\end{equation}
where \(\mu_\gamma(k)\) is the least Poisson mean attaining upper-tail
probability \(\gamma\). The implementation enumerates the integer
discontinuities of \(K/\ell_M\) through 512 charged orders. Indeed,
\(K/\ell_M\) has its unique minimum at
\(b_x=z/[2\operatorname{atanh}(2w_N/z)]\), while \(\lambda_M(b)\) increases;
within each integer cell, its right root therefore minimizes
\(\mu_\gamma(k)/\lambda_M(b)\). The excluded region
is certified using \(\mu_\gamma(513)/\lambda_M^\infty\). Even at \(\gamma=0.50\),
these lower bounds are 22.49 for SCF and 14.88 for uniform, far above the
incumbent thresholds. The resulting values are in
\cref{tab:poisson-thresholds}.

\begin{table}[t]
\centering
\footnotesize
\setlength{\tabcolsep}{3.1pt}
\begin{tabular}{rrrr}
\toprule
\(\gamma\) & SCF \(m^\star\) & Uniform \(m^\star\) & Reduction \\
\midrule
0.50 & 1.48195 & 1.93460 & 23.40\% \\
0.75 & 1.70165 & 2.13404 & 20.26\% \\
0.90 & 1.91531 & 2.32479 & 17.61\% \\
0.95 & 2.04940 & 2.44362 & 16.13\% \\
0.99 & 2.31707 & 2.67570 & 13.40\% \\
\bottomrule
\end{tabular}
\caption{Exact Poisson chance-recovery thresholds. Each mechanism separately
chooses \(b\) before realized arrivals.}
\label{tab:poisson-thresholds}
\end{table}

For the common-shock stress test, let
\(U\sim\operatorname{Gamma}(k,k)\) be drawn once per recovery window and
replace every conditional charged-order intensity by \(mU\lambda_M(b)\).
Marginally, the charged count is negative binomial with mean
\(m\lambda_M(b)\) and variance
\(m\lambda_M(b)+m^2\lambda_M(b)^2/k\). We invert its exact recurrence and use
the same integer-cell enumeration and excluded-region certificate as in the
Poisson calculation. Results for the 95\% target are in
\cref{tab:common-shock-thresholds}.

\begin{table}[t]
\centering
\scriptsize
\setlength{\tabcolsep}{2.2pt}
\begin{tabular}{lrrr}
\toprule
Arrival model & SCF \(m^\star\) & Uniform \(m^\star\) & Reduction \\
\midrule
Independent Poisson & 2.04940 & 2.44362 & 16.13\% \\
Gamma--Poisson, \(k=10\) & 2.97057 & 3.72561 & 20.27\% \\
Gamma--Poisson, \(k=2\) & 8.57941 & 11.04360 & 22.31\% \\
\bottomrule
\end{tabular}
\caption{Exact 95\% chance-recovery thresholds under a shared multiplicative
arrival shock. Smaller \(k\) means stronger overdispersion.}
\label{tab:common-shock-thresholds}
\end{table}

The Poisson assumption also makes the screened informed-order counts
distributionally comparable. SCF screens
\(\operatorname{Pois}(mi\ell_D)=\operatorname{Pois}(m\delta a)\), while the
sum across the two uniformly charged directions is
\(\operatorname{Pois}(2mi\ell_U)=\operatorname{Pois}(m\delta a)\). This is
equality in marginal distribution, not pathwise equality.

For the distributional stress test, let \(g_{\alpha,\beta}\) be a Beta density
on \([0,1]\) and replace the raw intensity by
\(m c_{T,s}g_{\alpha_T,\beta_T}(W/w_T)dW\). Its integral remains
\(m c_{T,s}w_T\), and Beta\((1,1)\) reproduces the main model. If \(F_T\) is
the associated CDF, the executed mass per unit \(m\) is
\begin{equation}
 c_{T,s}w_T\left[
  1-F_T\!\left(\frac{D(b)+\ell_s}{w_T}\right)
 \right],
 \label{eq:beta-execution}
\end{equation}
and the informed loss in direction \(s\) is
\begin{equation}
 iw_I\left[
 F_I\!\left(\frac{D(b)+\ell_s}{w_I}\right)
 -F_I\!\left(\frac{D(b)}{w_I}\right)
 \right].
 \label{eq:beta-loss}
\end{equation}
The optimizer allows the common loss budget to be slack, refines every
grid-detected one-dimensional local maximum after reparameterizing levies by
informed loss, and then searches liquidity on a tail-certified finite log
interval. A production
grid of 65 inner by 193 outer points is checked against 129 by 385 points; the
largest absolute change in \(R\) is \(4.89\times10^{-15}\).

\begin{table}[t]
\centering
\footnotesize
\setlength{\tabcolsep}{3.2pt}
\begin{tabular}{lrrrr}
\toprule
& \multicolumn{4}{c}{Noise-surplus shape} \\
Informed shape & L & U & C & H \\
\midrule
L & 18.489 & 32.384 & 31.077 & 38.345 \\
U &  6.463 & 22.836 & 21.874 & 30.436 \\
C &  3.133 & 16.535 & 16.249 & 25.504 \\
H &  0.203 &  0.689 &  0.030 &  3.748 \\
\bottomrule
\end{tabular}
\caption{Threshold reduction \(R\) (percent) outside the theorem under Beta
surplus shapes. L, U, C, and H denote Beta\((1,3)\), Beta\((1,1)\),
Beta\((2,2)\), and Beta\((3,1)\), respectively.}
\label{tab:beta-shapes}
\end{table}

The outputs are formula checks and synthetic stress tests, not observations from
a live market.

\end{document}